\documentclass[%
reprint,
 amsmath,amssymb,
 aps,
prb,
]{revtex4-1}

\usepackage{graphicx}
\usepackage{dcolumn}
\usepackage{bm}
\usepackage{color}
\usepackage{txfonts}
\usepackage{color}

\newcommand{\bvec}[1]{\mbox{\boldmath$#1$}}
\definecolor{green}{rgb}{0,0.75,0}

\definecolor{orange2}{rgb}{1,0.27,0}
\definecolor{indigo}{rgb}{0.5,0,0.7}

\begin{document}

\title{Strongly electron-correlated semimetal RuI$_3$ with a layered honeycomb structure}

\author{Kazuhiro Nawa$^1$}
\email{knawa@tohoku.ac.jp}
\author{Yoshinori Imai$^2$}
\author{Youhei Yamaji$^3$}
\author{Hideyuki Fujihara$^2$}
\author{Wakana Yamada$^2$}
\author{Ryotaro Takahashi$^1$}
\author{Takumi Hiraoka$^{1,2}$}
\author{Masato Hagihala$^4$}
\author{Shuki Torii$^4$}
\author{Takuya Aoyama$^2$}
\author{Takamasa Ohashi$^5$}
\author{Yasuhiro Shimizu$^5$}
\author{Hirotada Gotou$^6$}
\author{Masayuki Itoh$^5$}
\author{Kenya Ohgushi$^2$}
\author{Taku J. Sato$^1$}

\affiliation{
$^{1}$Institute of Multidisciplinary Research for Advanced Materials, Tohoku University, 2-1-1 Katahira, Sendai 980-8577, Japan \\
$^{2}$Department of Physics, Graduate School of Science, Tohoku University, 6-3 Aramaki-Aoba, Aoba-ku, Sendai, Miyagi 980-8578, Japan \\
$^{3}$Center for Green Research on Energy and Environmental Materials, National Institute for Materials Science, Namiki, Tsukuba-shi, Ibaraki, 305-0044, Japan \\
$^{4}$Institute of Materials Structure Science, High Energy Accelerator Research Organization (KEK), 203-1 Tokai, Ibaraki, 319-1106, Japan \\
$^{5}$Department of Physics, Graduate School of Science, Nagoya University, Furo-cho, Chikusa-ku, Nagoya 464-8602, Japan \\
$^{6}$Institute for Solid State Physics, University of Tokyo, Kashiwa, Chiba 277-8581, Japan
}

\begin{abstract}
A polymorph of RuI$_3$ synthesized under high pressure was found to have a two-layered honeycomb structure.
The resistivity of RuI$_3$ exhibits a semimetallic behavior, in contrast to insulating properties in $\alpha$-RuCl$_3$.
In addition, Pauli paramagnetic behavior was observed in the temperature dependence of a magnetic susceptibility and a nuclear spin-lattice relaxation rate 1/$T_1$.
The band structure calculations indicate that contribution of the I 5$p$ components to the low-energy $t_\mathrm{2g}$ bands effectively decreases Coulomb repulsion,
leading to semimetallic properties.
The physical properties also suggest strong electron correlations in RuI$_3$.
\end{abstract}


\maketitle

Interplay between a spin-orbit coupling (SOC) and electron correlations triggers a variety of intriguing electronic phases in 4$d$ and 5$d$ transition metal compounds\cite{SOC1, SOC2, SOC3}.
A large Coulomb repulsion makes spin-orbit coupled pseudospin degrees of freedom localized on single atoms, leading to a spin-orbit coupled Mott insulating state. 
Bond-dependent anisotropic magnetic interactions between pseudospin degrees of freedom stabilize a Kitaev spin liquid state\cite{Kitaev, JKmodel}.
On the other hand, if a Coulomb repulsion is small enough to recover a metallic behavior, 
hybridization induced by SOC results in a topologically insulating/semimetallic state\cite{TI}.
To investigate the interplay between two effects, it is required to tune an electronic state by controlling a band gap, 
from a spin-orbit coupled Mott insulating state to a topologically insulating/semimetallic state\cite{TopologicalSemimental, TopologicalSemimental2}.. 
Only limited compounds\cite{R2Ir2O7, R2Ir2O7_2, TopologicalSemimental, TopologicalSemimental2} allow the band-gap control to date.

Among a number of spin-orbit coupled Mott insulators, RuCl$_3$\cite{RuCl3_el2, RuCl3_el3} has attracted focused attention due to the Kitaev spin liquid behavior.
Ru$^{3+}$(4$d^5$) carries pseudospin 1/2, and forms a layered honeycomb structure.
A few polymorphs, which crystallize into the space group of $C2/m$\cite{RuCl3_1, RuCl3_2, RuCl3_4}, $P3_121$\cite{RuCl3_0}, and $P3_1$\cite{RuCl3_5}, have been discovered.
A low-temperature structure of $R\overline{3}$ is also indicated from single crystalline XRD\cite{RuCl3_3}, Raman\cite{Raman}, and NMR experiments\cite{NMR}.
These polymorphs have different stacking sequences whereas all of them form a layered honeycomb structure.
RuCl$_3$ exhibits a zigzag antiferromagnetic order at 8-14~K\cite{zigzag, RuCl3_1, RuCl3_2}.
In spite of the presence of the magnetic order,
the majority of the spin excitations forms magnetic continuum that remains above the transition temperature.
These excitations were interpreted as the signature of a short-range order of the Kitaev spin liquid\cite{RuCl3_6, RuCl3_7}.

One approach to tune SOC and electron correlations is to substitute Cl atoms and form Ru$X_3$ ($X$ = Br, and I).
The effect of halogen substitution has been only investigated though density functional theory (DFT) calculations\cite{DFT1, DFT2}.
Very recently, high pressure syntheses of their polymorphs have been reported\cite{RuBr3, RuI3}.
A polymorph of RuBr$_3$ is found to have a three-layered honeycomb structure with the space group of $R\overline{3}$,
which is quite close to the low-temperature form of RuCl$_3$\cite{RuBr3}.
It is understood as a spin-orbit coupled Mott insulator, since the resistivity follows a thermally activated temperature dependence similar to RuCl$_3$\cite{RuCl3_el1, yi19}.
At high temperatures, the magnetic susceptibility well follows the Curie-Weiss rule, indicating localization of pseudospin degrees of freedom.
In this paper, we report the synthesis of a polymorph of RuI$_3$, which also has a layered honeycomb structure.
The resistivity of RuI$_3$ exhibits a semimetallic behavior.
Pauli paramagnetic behavior was observed in a magnetic susceptibility and a \textcolor{black}{nuclear} spin-lattice relaxation rate 1/$T_1$.
These physical properties are different from those observed in $\alpha$-RuCl$_3$ and RuBr$_3$.
The halogen substitution largely changes the electronic structure in Ru$X_3$.

Polycrystalline samples of RuI$_3$ with a layered honeycomb structure were synthesized by using a cubic-anvil-type high-pressure apparatus.
A starting material of commercially-available RuI$_3$ was placed in a platinum capsule, and loaded into a pyrophyllite cube.
Then it was pressurized at 4~GPa and heated at 400$^\circ$C for 30 minutes.

Powder X-ray diffraction  (PXRD) patterns were collected at room temperature using in-house rotating anode X-ray generator (Mo K$_\alpha$, Rigaku RA MultiMax9)
equipped with a Eulerian cradle (Huber).
The polycrystalline samples were sealed in soda-lime glass capillaries with the diameter of 0.3~mm under Ar atmosphere, and intensities were collected with spinning the capillaries.
Powder neutron diffraction (PND) experiments were performed using the high-resolution TOF neutron powder diffractometer SuperHRPD installed at J-PARC\cite{SHRPD}.
The powder sample of 1.54 g was loaded into a vanadium-nickel cylindrical sample can with a diameter of 6~mm.
The sample can was then set to the $^4$He closed-cycle refrigerator with the lowest attainable temperature of 3 K.
Rietveld analyses were made using the Fullprof software suite\cite{Fullprof}. 
The pseudo-Voigt function convoluted with back-to back exponentials\cite{BBEF} was used to fit the peak profile.
The phenomenological model of anisotropic peak broadening induced from strain was applied to the peak width\cite{APB}.

The electrical transport properties up to 9~T including the resistivity, magnetoresistance and Hall resistivity were measured by the standard four-terminal method with a use of a commercial apparatus (PPMS: Physical Property Measurement System, Quantum Design).
The heat capacity was measured by the thermal-relaxation method down to 2 K by using PPMS. 
Magnetic susceptibility measurements were performed using a superconducting quantum interference device (SQUID) magnetometer.
The $^{127}$I \textcolor{black}{nuclear quadrupole resonance (NQR)} measurement was performed by using the spin echo technique.
The spin echo signal was obtained after the $\pi$/2 pulses with the length of 1.5~$\mu$s and the interval time $\tau$ = 10~$\mu$s.
The nuclear spin-lattice relaxation rate $1/T_1$ was measured by the saturation recovery method at zero field. 

\begin{figure}[t]
\centering
\includegraphics[width=0.8\linewidth]{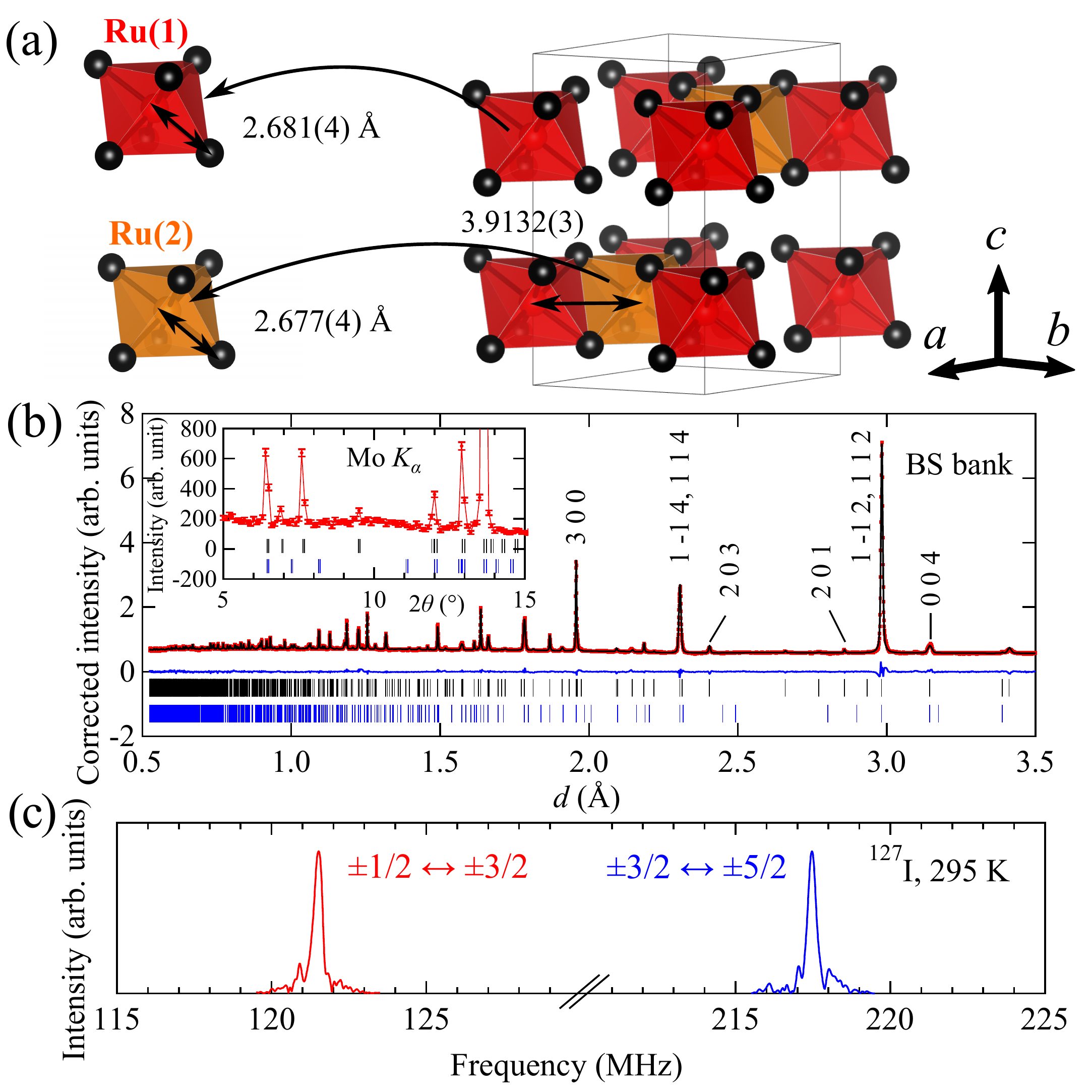}
\caption{\label{structure} (a) Crystal structure of RuI$_3$ with a two layered honeycomb structure. Ru(1), Ru(2), and I atoms are colored by red, orange, and black, respectively.
(b) PND pattern collected at 300~K by the BS bank of the SuperHRPD diffractometer. Red dots, black, and blue curves represent the observed,
calculated intensities, and the difference between them.
The inset shows a PXRD pattern at low angles. In both figures, the expected peak positions of the two- ($P\overline{3}1c$, black) and three-layered ($R\overline{3}$, blue) honeycomb structures are plotted. (c) $^{127}$I NQR spectrum measured at zero field and 295 K. }
\end{figure}

PXRD and PND patterns indicate a two-layered honeycomb structure
for the polymorph of RuI$_3$, as shown in Fig.~\ref{structure}(a).
The PND and PXRD patterns are shown in the main figure and the inset of Fig.\ref{structure}(b), respectively.
The PXRD pattern of the post-sintered sample is different from the pattern of the starting material that reflects the one-dimensional structure\cite{betaRuI3, betaRuI3_2} (see Supplemental Information for details\cite{SI}).
The peak position of both patterns cannot be explained by a three-layered honeycomb structure with the space group of $C2/c$\cite{RuCl3_2, RuCl3_4} and $R\overline{3}$\cite{RuCl3_3, RuBr3}.
The difference in the extinction rule indicates a significant change in the stacking sequence.

Transition-metal trihalides form a variety of layered crystal structures\cite{AX3_0, AX3_1, AX3_2, AX3_3}.
Among them, a structure found in a polymorph of TiCl$_3$ well reproduces the peak positions of the two patterns, as indicated by black vertical bars in Fig.~\ref{structure}(b). 
The PND patterns were well fitted by the two-layered honeycomb structure model with the reasonable $R$-factors.
The structure is represented by the space group $P\overline{3}$1$c$ with lattice constants $a =~6.7779(4)$ and $c =~12.5794(23)~\AA$ at room temperature. 
Details of the structure parameters and the refined crystallographic data are listed in Table~S1 in the Supplemental Information\cite{SI}.
In RuI$_3$ synthesized under high pressure, I atoms form the hexagonal closed-packed structure.
The stacking sequence of halogen atoms is the same as that in RuBr$_3$\cite{RuBr3}, while it is different from that in $\alpha$-RuCl$_3$,
where Cl atoms form the cubic closed-packed structure. 
Ru atoms occupy two-third of vacancies for every two I layers, resulting in the formation of a honeycomb lattice inside the $ab$-plane.
The unit cell includes two layers of the Ru honeycomb plane. In each plane, Ru atoms occupy two inequivalent positions: Ru(1) atoms at 2$a$ site and Ru(2) atoms at 2$d$ site.
Since a threefold rotation axis along $c$-direction is present at the both Ru sites, the network formed by nearest neighbor Ru--Ru bonds can be regarded as a regular honeycomb lattice.
The difference in the two Ru atoms is their stacking sequence along the $c$-axis:
Ru(1) atoms stack uniformly along the $c$ direction while Ru(2)  atoms stack in a staggered manner.
The difference is due to inversion symmetry that exists at the center of the two Ru(1) atoms on neighboring planes.
The distance of the nearest neighbor Ru(1)--Ru(2) bond, a Ru(1)--I bond, and a Ru(2)--I bond are \textcolor{black}{3.9132(3)}, \textcolor{black}{2.681(4)}, and \textcolor{black}{2.677(4)} \AA, respectively.
The bond angle of the the nearest neighbor Ru(1)--I--Ru(2) bond is \textcolor{black}{93.85(8)}$^\circ$ at 300~K.
The PND patterns were well fit by the space group of $P\overline{3}$1$c$ down to 3~K, and no signature of the structural transition such as a split of peaks was found\cite{SI}.

$^{127}$I NQR spectra do not contradict with the local symmetry of I atoms expected from the neutron diffraction experiments. 
Figure~\ref{structure}(c) shows the $^{127}$I ($I$ = 5/2, $Q$ = 0.721$b$) \textcolor{black}{NQR} spectrum at 295 K.
The sharp resonance lines at 121.5 and 217.5 MHz come from \textcolor{black}{$\pm1/2 \leftrightarrow~\pm3/2$ and $\pm3/2~\leftrightarrow~\pm5/2$ transitions}, respectively.
Only a single line is found for each resonance, which is consistent with one I site expected from $P\overline{3}$1$c$.
From the NQR frequencies, we could uniquely determine the asymmetry of electric field gradient $\eta$ = $|V_{xx} - V_{yy}/V_{zz}|$ = 0.32.
No signature of magnetic and structural phase transition is found down to 4~K, which is also consistent with the PND experiments\cite{SI}.

\begin{figure}[t]
\begin{center} 
\includegraphics[width=0.8\linewidth]{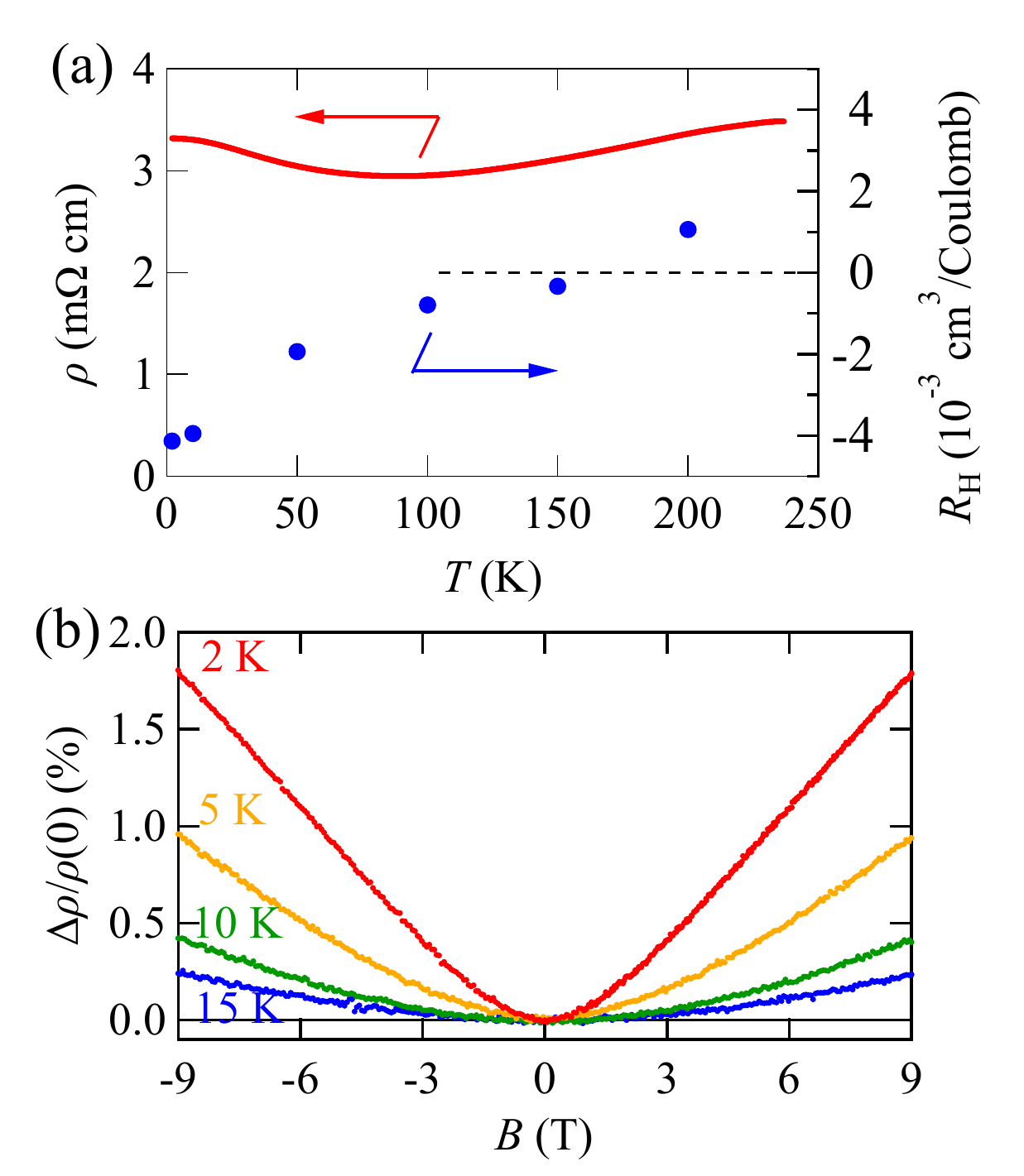}
\vspace{-5mm}
\caption{
(a) Temperature dependence of a zero-field resistivity $\rho$ and a Hall coefficient $R_\mathrm{H}$.
(b) Magnetoresistance up to $\pm$9~T, $\Delta \rho/\rho(0) = (\rho(B) - \rho(0))/\rho(0)$. The data are taken in a transverse geometry.
}
\label{fig:rho}
\end{center}
\vspace{-5mm}
\end{figure}

Electronic properties of RuI$_3$ can be explained by a semimetal with multiple carriers.
Figure \ref{fig:rho}(a) shows a temperature dependence of a zero-field resistivity $\rho$ for RuI$_3$.
On cooling from room temperature, the resistivity gradually decreases, shows a small upturn at 90 K, and nearly saturates at low temperature.
The temperature variation is small in the whole measured temperature range, and the value of $\rho \sim 3-3.5$ m$\Omega$cm is rather large as compared to a simple metal.
These results infer that RuI$_3$ is a semimetal and/or a correlated metal.
The metallic feature of RuI$_3$ is in stark contrast to an insulating behavior in $\alpha$-RuCl$_3$ and RuBr$_3$\cite{yi19, RuBr3}, indicating that the electronic state of RuI$_3$ should be totally different from that of $\alpha$-RuCl$_3$ and RuBr$_3$.

Figure \ref{fig:rho}(b) shows a magnetoresistance (MR), $\Delta \rho / \rho(0) = \left ( \rho \left ( B \right ) - \rho \left ( 0 \right ) \right )/\rho \left ( 0 \right )$, at 2--20~K, where a magnetic field $B$ was applied perpendicular to the current direction. 
The positive MR following a $B^2$ dependence was observed below 20 K, which can be interpreted as a normal MR in a multiple-carrier system.
The rapid evolution of a positive MR on cooling indicates the large enhancement of carrier mobilities.
This seemingly contradicts the observed negligible temperature dependence of $\rho$.
To explain the temperature variation of $\rho$, one should consider the possibilities that the carrier densities and/or quasiparticle mass depend on temperature in a complicated manner.
One of possible scenarios is a generation of thermally-excited carriers at elevated temperatures in a semimetal.
At the base temperature of 2~K, MR shows an unusual linear $B$ dependence in the high magnetic field regime, $B > \sim 2.3$ T.
Such a feature is frequently observed in a semimetal with a linear band dispersion such as Dirac and Weyl semimetals \cite{DiracConeStates,PRL.115.027006}.
Although our band calculations, which are discussed later, also indicate the presence of the linear band dispersion,
further calculations are necessary to confirm its existence near the Fermi surface.

A temperature dependence of a Hall coefficient $R_\mathrm{H}$ indicates coexistence of hole and electron carriers.
The Hall resistivity \textcolor{black}{$\rho_\mathrm{H}$} almost linearly depends on $B$ over the entire magnetic field range, and
$R_\mathrm{H}$ is estimated from a linear fit. 
As shown in Fig. \ref{fig:rho}(a), $R_\mathrm{H}$ is positive at 200 K and decreases in magnitude with decreasing temperature;
it then shows a sign change at 150 K and increases in magnitude with further decreasing temperature.
This suggests that dominant carriers change from holes at high temperature to electrons at low temperature.
In a framework of a single carrier model, the carrier density at 2 K is estimated to be $1.51 \times 10^{21}$ $\mathrm{cm}^{-3}$ (0.873 per a single unit cell); one has to apply a more sophisticated model for the further understanding (see Supplemental Information for further discussion\cite{SI}).

\begin{figure}[t]
\begin{center}
\includegraphics[width=0.8\linewidth]{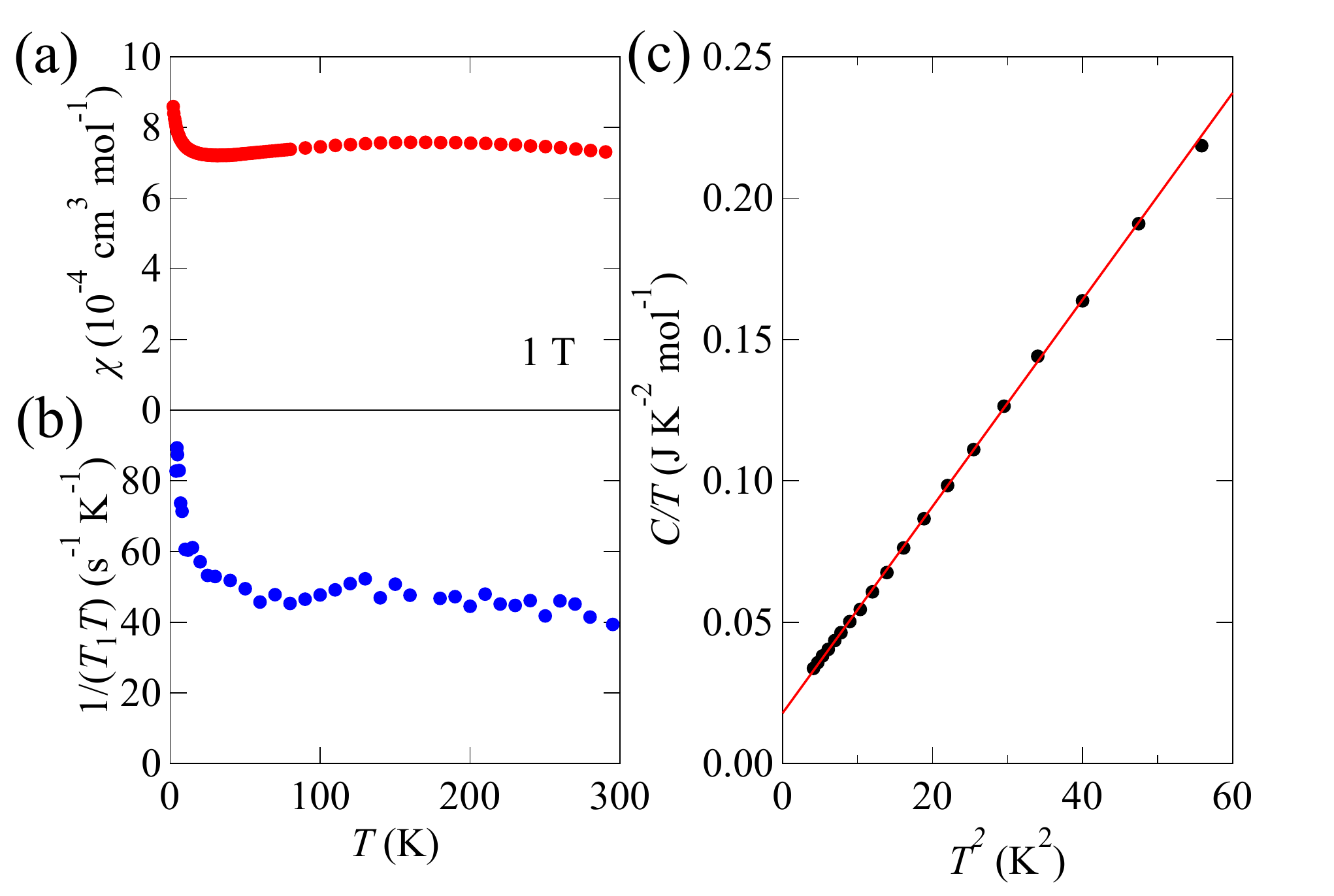}
\caption{
Temperature dependence of (a) a magnetic susceptibility $\chi$ measured at 1 T and (b) $^{127}$I nuclear spin-lattice relaxation rate $1/T_1$ at a zero magnetic field.
(c) The heat capacity divided by the temperature $C/T$ plotted as a function of the squared temperature. The solid red line represents a fit to a linear function.
}
\label{fig:chi}
\end{center}
\vspace{-5mm}
\end{figure}

Thermodynamic properties of RuI$_3$ are characterized by Pauli paramagnetic features. 
Figure \ref{fig:chi}(a) represents the temperature $T$ dependence of magnetic susceptibility, $\chi$, at $B=1$~T.
One cannot find any signatures of a magnetic transition at 2--300 K.
A temperature dependence of the magnetic susceptibility is small in a wide temperature range, which indicates a Pauli paramagnetic state.
Below 150 K, one can barely observe a suppression of $\chi$, which is attributable to the pseudo-gap structure in the density of states.
An increase below 20 K is most likely due to the Curie term from a small number of impurities. 
The $\chi$ data is fitted by the sum of a Pauli paramagnetic term, $\chi_0$, and a Curie term: $\chi = \chi_0+C_\mathrm{CW}/T$, where $C_\mathrm{CW}$ is a Curie constant.
The best fit yields $\chi_0=7.11 \times 10^{-4}$ emu/mol and $C_\mathrm{CW}=1.14 \times 10^{-3}$ K cm$^{3}$/mol.

Figure \ref{fig:chi}(b) shows a temperature dependence of the nuclear spin-lattice relaxation rate divided by $T$, 1/$(T_1T)$, measured for the \textcolor{black}{$\pm1/2~\leftrightarrow~\pm3/2$ transition}.
The nearly temperature independent behavior at high temperatures is consistent with the Pauli paramagnetic behavior observed in the magnetic susceptibility.
An increase in 1/$(T_1T)$ was observed below 40 K.
This suggests the development of antiferromagnetic correlation as expected for a metallic state with strong electron correlations.

The heat capacity divided by the temperature, $C/T$, under a zero magnetic field is plotted against squared temperature, $T^2$, in Fig.~\ref{fig:chi}(c).
The data is well fit by a sum of the electron and phonon components: $C = \gamma T + \beta T^3$.
The fitting curve is shown by the solid line in Fig.~~\ref{fig:chi}(c).
The fit yields $\gamma=17.7$ mJ/K$^2$ mol and $\beta=3.66$ mJ/K$^4$ mol. 
The Wilson ratio, $R_\mathrm{w}=\pi^2k_\mathrm{B}^2\chi_0/3\mu_\mathrm{B}^2\gamma$ \textcolor{black}{($k_\mathrm{B}$ being Boltzmann constant)}, is estimated to be 2.93, indicating a strong electron correlation in Ru 4$d$ bands.
The Debye temperature is estimated as $\theta_\mathrm{D} = 129$ K from the formula $\beta=12\pi^4NR/5\theta_\mathrm{D}^3$, where $N=4$ is the number of atoms in a formula unit and $R$ is the gas constant.
In summary, electronic properties show that RuI$_3$ is a strongly-electron-correlated semimetal,
of which ground state may be located in the vicinity of the Mott transition.

To elucidate the origin of semimetallic properties, electronic structures of RuI$_3$ were investigated by denisty-functional theory calculations.
The band structure for RuI$_3$ is shown in Figs.~\ref{Fig_alpha-RuI3_band}(a) and \ref{Fig_alpha-RuI3_band}(b). 
The calculations were performed by using Q{\sc uantum} ESPRESSO~\cite{Giannozzi_2009,Giannozzi_2017} based on the experimental crystal structure at 300 K.  
Fully relativistic calculations was performed with a generalized gradient approximation (GGA) exchange-correlation functional.

Chemical trend of halogen substitution was investigated by comparing the electronic structure of RuI$_3$ with $P\overline{3}1$c
with those of $\alpha$-RuCl$_3$, RuBr$_3$ and RuI$_3$ with the $R\overline{3}$ structure.
As shown in Fig.~\ref{Fig_alpha-RuI3_band}(a), when the atomic number of the halogen increases from Cl 3$p$ to Br 4$p$ and to I 5$p$,
the $p$ orbital components of the halogen atoms become closer to the Fermi level, irrespective of the difference between $R\overline{3}$ and $P\overline{3}1$c.
The halogen $p$ components in the low-energy $t_{2g}$ bands around the Fermi level become larger with increasing the atomic number.
The Wannier orbitals corresponding to the $t_{2g}$ bands are more extended in RuI$_3$, 
while the $t_{2g}$ band width remains narrow and similar in these three Ru halides.
The spread of the $t_{2g}$ Wannier orbitals at Ru(1) (Ru(2)) of RuI$_3$ crystalline in $P\overline{3}1c$ is 7.04 $\AA^2$ (6.90 $\AA^2$),
which is larger than the spread of RuBr$_3$ and RuCl$_3$ in the $R\overline{3}$ structure, 5.26 $\AA^2$ and 3.86 $\AA^2$, respectively.
The shallower halogen $p$ levels introduce stronger screening of the Coulomb repulsion in the $t_{2g}$ bands.
In more accurate treatments of electron correlations beyond GGA,
Coulomb repulsion $U$ should induce a band gap and make the ground state a spin-orbit coupled Mott insulating state.
Increase in the hybridization between the Ru $d$ orbitals and halogen $p$ orbitals plausibily decreases the effective intra-atomic Coulomb repulsion $U$ in the low-energy $t_{2g}$ manifold, and, thus, reduces the band gap.
This naturally explains the metallic behavior in RuI$_3$, in contrast to insulating behavior found in $\alpha$-RuCl$_3$ and RuBr$_3$.

The present relativistic GGA band structures shown in Fig.~\ref{Fig_alpha-RuI3_band}(a) indicates that RuI$_3$ is a compensated semimetal.
The Fermi surface consists of three (doubly-degenerated) hourglass-shaped hole surfaces centered at the M points and
two (doubly-degenerated) cylindrical electron surfaces, as shown in Fig.~\ref{Fig_alpha-RuI3_band}(c).
The coexistence of the hole and electron surfaces is consistent with the sign change observed in the Hall coefficient.
Carrier densities are estimated as \textcolor{black}{0.188 hole and 0.188 electron carriers per single unit cell}.
The large deviation from that estimated from the experiment reflects the presence of several carriers contributing to Hall coefficients~\cite{SI}.

The energy scales of magnetic exchange in RuI$_3$ were analyzed by constructing maxmally localized Wannier orbitals~\cite{marzari2012maximally,Pizzi2020} for the low-energy $t_{2g}$ bands.
(See Supplemental Materials for the details\cite{SI}).
From the $ab$ $initio$ Wannier analysis,
the on-site SOC is represented by
$\zeta_{\rm so}^{\parallel}\bvec{s}_{\parallel}\cdot\bvec{\ell}_{\parallel} + \zeta_{\rm so}^{\perp}s_{\perp}\ell_{\perp}$.
Here, $\bvec{s}_{\parallel}$ and $s_{\perp}$ ($\bvec{\ell}_{\parallel}$ and $\ell_{\perp}$) are the in-plane
and out-of-plane ($c$ axis) components of spins (effective angular momenta), respectively.
{The SOCs are estimated as $\zeta_{\rm so}^{\parallel} = $ 0.188~eV (0.187~eV) and $\zeta_{\rm so}^{\perp} = $ 0.134~eV (0.136~eV)
for the Ru(1) (Ru(2)) sublattice.
Because of the contribution of I 5$p$ atomic orbitals to the Wannier orbitals,
the SOCs become larger than RuBr$_3$ in the $R\overline{3}$ structure as $\zeta_{\rm so}^{\parallel}=$ 0.13~eV and
$\zeta_{\rm so}^{\perp}=$ 0.12~eV (see also Ref.~\cite{PhysRevB.100.075110} for RuCl$_3$ in the $C2/m$).
The inequivalence of Ru(1) and Ru(2) introduces different 
on-site potentials of $\epsilon - E_{\rm F}=-$0.420~meV and $-$0.410~meV for Ru(1) and Ru(2), respectively.
The largest nearest-neighbor hopping, which is relevant to the generation of the Kitaev coupling, is 0.162~eV.
The largest inter-layer hopping between two adjacent RuI$_3$ layers is also as large as $-$0.049~eV. 
Since strong SOC would induce topological natures in the band structure, 
further band structure calculations would reveal interesting nodal semimetallic states in RuI$_3$. 

\begin{figure}[t]
\begin{center}
\includegraphics[width=0.5\textwidth]{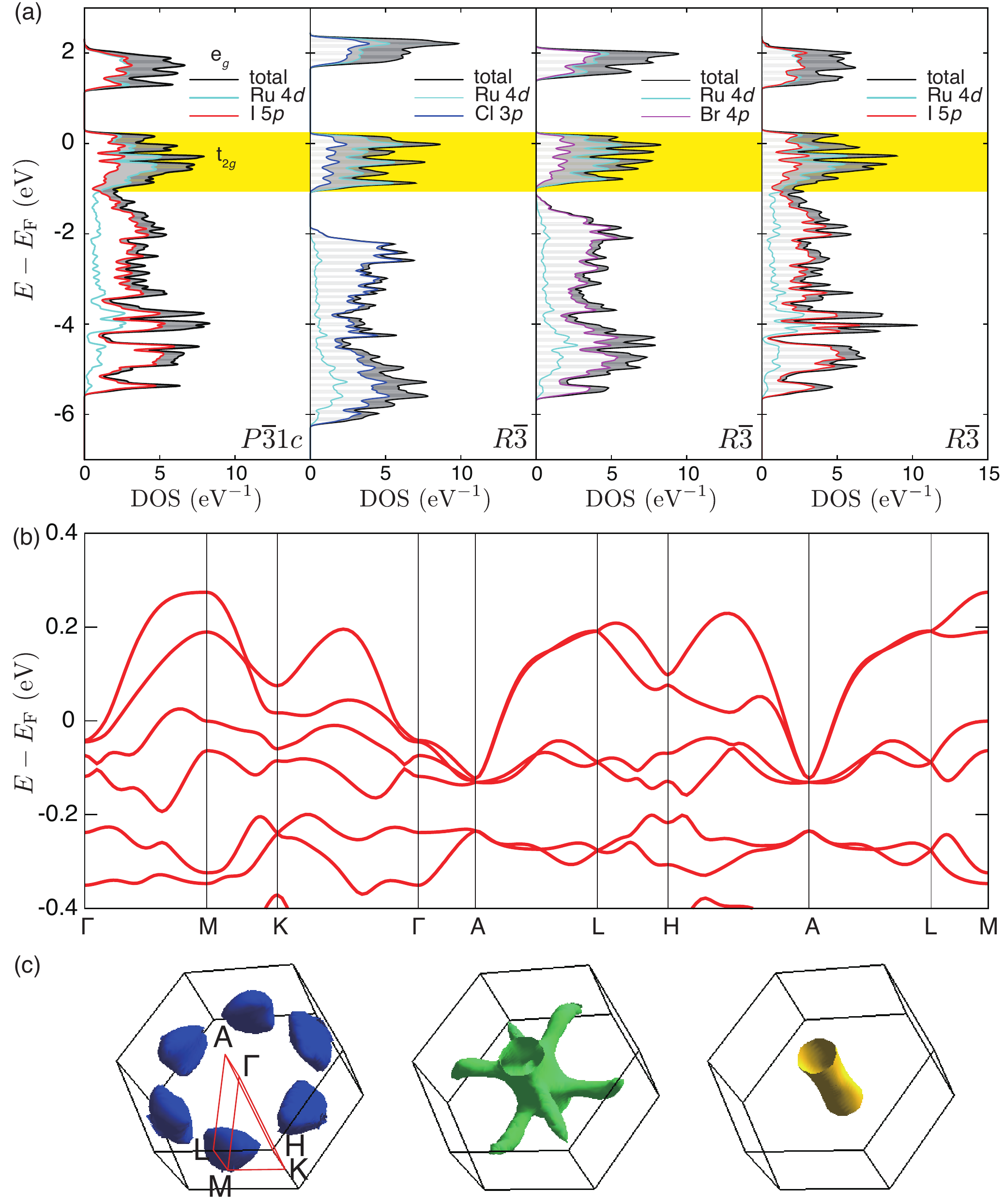}
\end{center}
\caption{
Electronic structure
and Fermi surfaces of RuI$_3$ by GGA+SOC calculations. 
(a) The density of states (DOS) of $\alpha$-RuI$_3$ in the $P\overline{3}1c$ structure (the left end) and Ru$X_3$ ($X$=Cl, Br, and I)
with the $R\overline{3}$ structure are shown by black solid curves.
The partial density of states projected onto atomic Ru 4$d$, Cl 3$p$, Br 4$p$, and I 5$p$ orbitals are also shown
by cyan, blue, magenta, and red curves, respectively.
The DOS of the $t_{2g}$ bands is highlighted by the yellow belt.
(b) The band dispersion mainly consist of 4$d$ orbitals of Ru
and 5$p$ orbitals of I atoms is shown along the symmetry lines (see also (c)) by focusing on the 4 doubly-degenerated low-energy bands in the $t_{2g}$ manifold.
(c) The Fermi surfaces are illustrated~\cite{KAWAMURA2019197}
in the first Brillouin zone.
From the left to right,
three hourglass-shape hole pockets centered at the M point (left), quasi-two dimensional electron surfaces (middle),
and cylinderical electron surfaces (right) are shown. 
}
\label{Fig_alpha-RuI3_band}
\end{figure}

In summary, we report the crystal structure and the physical properties of a polymorph of RuI$_3$, which is made by high pressure synthesis.
PXRD and PND patterns reveal that the polymorph crystallizes into $P\overline{3}1c$ with a two-layered honeycomb structure.
The electronic properties indicate a semimetallic state composed by multiple carriers, which is supported by the density functional theory calculations.
Magnetic properties also support delocalization of pseudospin degrees of freedom.
Hybridization of I 5$p$ orbitals with the low-energy $t_{2g}$ bands should tune RuI$_3$ to a semimetal with strong electron correlations.

Note: We find the manuscript reporting crystal structure and physical properties of the polymorph of RuI$_3$ during preparation of the manuscript\cite{RuI3}.
The manuscript reports a three-layered honeycomb structure (space group $R\overline{3}$), which is different from the two-layered honeycomb structure ($P\overline{3}1c$) discussed in this paper. 
It should be stressed that our PXRD and PND patters cannot be indexed by the space group $R\overline{3}$, which indicates that the two different polymorphs are synthesized
at distict pressure and temperature conditions.
On the other hand, physical properties are similar, such as a semimetallic behavior found in electrical resistivity. 

\textbf{Acknowledgements}
We would like to thank Takeshi Yajima for his helps in the X-ray diffraction measurements.
This work was carried out under the Visiting Researcher's Program of the Institute for Solid State Physics, the University of Tokyo.
\textcolor{black}{The neutron diffraction experiments at the Materials and Life Science Experimental Facility of the J-PARC were performed under a user program (Proposal No. 2020B0430).}
The present work was financially supported by JSPS KAKENHI Grant Numbers JP18H01159, JP18K03531, JP19H01837, JP19H04685, JP19H05822, JP19H05823, JP19K21837 and JP20H01850, Murata Science Foundation, and the CORE Laboratory Research
Program “Dynamic Alliance for Open Innovation Bridging Human, Environment and Materials” of the Network Joint Research Center for Materials and Device.
This work was also supported by JST CREST Grant no. JP19198318, Japan.
Y. Y. was supported by by MEXT as ``Basic Science for Emergence and Functionality in Quantum Matter
- Innovative Strongly-Correlated Electron Science by Integration of Fugaku and Frontier Experiments -" as a program for promoting researches on the supercomputer Fugaku, 
supported by RIKEN-Center for Computational Science (R-CCS) through HPCI System Research Project (Project ID: hp210163).

\bibliographystyle{apsrev4-1}
\bibliography{exprep}

\end{document}